\documentclass[11pt,twoside,a4paper]{article}
\usepackage{amsfonts}
\usepackage{amsmath}
\usepackage{graphicx}
\usepackage{amsmath}
\usepackage[colorlinks]{hyperref}

\topmargin by -\headsep

\evensidemargin 
\oddsidemargin 
\advance\hoffset by -3mm  
\advance\voffset by  8mm  

\newtheorem{theorem}{Theorem}
\newtheorem{acknowledgement}{Acknowledgement}

\newtheorem{example}[theorem]{Example}

\newtheorem{remark}[subsection]{Remark}

\newenvironment{proof}[1][Proof]{\textbf{#1.} }{\ \rule{0.5em}{0.5em}}

\begin{document}

\title{Ordinary Differential Equations through Dimensional Analysis.}
\author{J.A. Belinch\'{o}n\thanks{%
E-mail:abelcal@ciccp.es} \\
{\small Dpt. F\'{\i}sica. ETS Arquitectura. Av. Juan de Herrera 4. Madrid
28040 Espa\~{n}a.}}
\date{{\footnotesize Last Draft (\today)}}
\maketitle

\begin{abstract}
In in this paper we show how using D.A. it is found a simple change of
variables (c.v.) that brings us to obtain differential equations simpler
than the original one. In a pedagogical way (at least we try to do that) and
in order to make see that each c.v. corresponds to an invariant solution
(induced by a symmetry) or a particular solution, we compare (with all the
tedious details, i.e. calculations) the proposed method with the Lie method.
The method is checked even in odes that do not admit symmetries.

\end{abstract}

\tableofcontents

\section{Introduction}

D.A. has usually been employed in different areas (fields) such as
engineering problems, fluid mechanics etc... and these problems are always
described by partial differential equations (pde) (see \cite{ad2}-\cite%
{Barenblat}). This method (\textquotedblleft tactic") helps us to reduce the
number of quantities that appear into an equation and to obtain ordinary
differential equations (ode). We would like to point out (emphasize) that
this tool is more effective if one practices the spatial discrimination,
such tactic allows us to obtain better results than with the standard
application of D.A. (see \cite{cas1}). Knowing that D.A. works well in pde
we would like to extend this method to the study of ode (the first order in
this case) in a systematic way. There are in the literature previous work in
this direction for ode of first order (see \cite{tony1}-\cite{Castans1}.

The method of the Lie groups it has showed as a very useful tool in order to
solve nonlinear equations (nl-ode) as well as pde (see \cite{1902}-\cite%
{Olver} ). Nevertheless, when one is studying ode of first order its
application results very complicated (very tedious) if one has not a
computer algebra package since if one decides to look for the possible
symmetries of an ode with pencil and paper this task may be turned very
exhausting. \ However, if we know that an ode admits a concrete symmetry,
then it is a trivial issue to find new variables which allows us to rewrite
the ode in quadratures or as we will see in this paper to obtain an ode with
separating variables.

Our purpose in this work is to explain, through examples, how D.A. works in
order to find these changes of variables (c.v.) in a trivial way, i.e.
without the knowledge of the symmetries of the ode under study. The idea is
as follows. When we are studying an ode form the dimensional point of view,
we must require that such ode verifies the principle of dimensional
homogeneity (pdh) i.e. that each term within the equation have the same
dimensions, for example, speed, or energy density. To clarify this concept
we consider the following ode, $y^{\prime }=\dfrac{y}{x}+x,$ where each term
must have dimensions of $y^{\prime }$ i.e. $\left[ y^{\prime }\right]
=yx^{-1},$ where $\left[ \cdot \right] $ stands for dimensional equation of
the quantity $\cdot .$ As we can see, this ode does not verify the pdh,
since the term $x$ has dimensions of $x,$ i.e. $\left[ x\right] =x.$ In
order to do that this equation verifies the pdh we need to introduce
dimensional constants, in such a way that after rewrite the ode with these
constants the ode now verifies such principle i.e. in this example and as we
can see easily if we consider only one constant $a$, such that $\left[ a%
\right] =yx^{-2}$, we make that the ode $y^{\prime }=\dfrac{y}{x}+ax,$
verifies the pdh. We would like to emphasize that this situation does not
appear (arise) when one is studying physical or engineering problems since
(as it is supposed) that such problems (equations) verify the pdh and we do
not need to introduce new dimensional constants, that must have physical
meaning, for example, the viscosity coefficient, etc....

Precisely this dimensional constant suggests us the c.v. $\left(
x,y(x)\right) \longmapsto \left( t,u(t)\right) $ where $\left(
t=x,u=yx^{-2}\right) $ in such a way that rewriting the original ode in
these new variables it is obtained a new ode with separating variables: $%
u+u^{\prime }t=1.$ The reason is the following. We know from the Lie group
theory that if it is known a symmetry of an ode then this symmetry bring us
through a c.v. to obtain a simpler ode (a quadrature or a ode with
separating variables) and therefore the solution is found in a closed form.
To find this c.v. it is used the invariants that generate each symmetry,
i.e. the first principle is that it is useful to pass to new coordinates
such that one of the coordinate functions is an invariant of the group.
After such transformation it often (but not always) happens that the
variables separate and the equation can be solved in closed form. We must
stress that taking the new dependent variables to be an invariant of the
group does not guarantee the separation of variables. The choice of the
independent variable is also very important.

One must note that D.A. gives us this invariant (or at least a particular
solution). \ It is observed in our example that the solution $y=x^{2}$,
(suggested by D.A.) is a particular solution for this ode, but if we study
this ode form the Lie theory point of view it is obtained that such ode has
the scaling symmetry $X=x\partial _{x}+2y\partial _{y}$ and therefore the
solution $y=x^{2}$, is furthermore an invariant solution (generated by $X$),
for this reason such c.v. works well. As we will show in the paper, the c.v.
that D.A. induces is the same than the generated one by the symmetries of
the ode. In order to make see this fact, we will solve each example by D.A.
as well as by the Lie group technique (abusing of the trivial calculations),
calculating the symmetries and their corresponding invariants and c.v..

However D.A. has a little (or rather big) drawback, it only gives us
relationship of type power i.e. $y=x^{n},$ with $n\in \mathbb{R},$ and we
cannot obtain relationship as $y=\left( x+1\right) ^{n}$. Nevertheless, and
as we will see in the examples, although D.A. does not give us an invariant
solution it will be sufficient that it provides us a particular solution in
order to obtain a c.v. which brings us to a simpler ode than the original
one. This task will be very useful in the case of studying Riccati and Abel
odes. Furthermore, this \textquotedblleft tactic" is valid even in the case
in which the ode has no symmetries but unfortunately one always founds
examples in which any tactic does not work.

The paper is organized as follows: In section 2 we describe (with all the
tedious and superfluous calculations), through two examples, how the
proposed method works by comparing it with the Lie method in order to show
which are the symmetries and their corresponding invariants. In section 3 we
will show several examples beginning with a Bernoulli ode and following with
two Riccati ode and three Abel ode (see \cite{kamke}-\cite{Abeles}). We
think that these kind of ode are the most difficult to solve and therefore
any simplification is welcome. In section 4 we show two Abel odes that do
not admit symmetries (see \cite{cheb}-\cite{Green}). In the first of them we
use the D.A. to obtain a particular solution that allows us to obtain a
Bernoulli ode but in the second one we are not able to find any particular
solution and therefore a c.v. which brings us to obtain a simpler ode. We
end with some conclusion as well as pointing out some of the limitations of
the proposed tactic.

We would like to emphasize the pedagogical character of this paper (at least
we try) for this reason we have abused of superfluous calculations and we
have omitted some technical details since the supposed audience are mainly
engineers and/or physicists but not for mathematicians.

\section{The method}

In this section we will explain how the D.A. works in order to solve odes \
through two simple examples. The main idea is to introduce dimensional
constants that make the ode under study verify the principle of dimensional
homogeneity. These dimensional constants help us to find c.v. which bring us
to obtain ode with separating variables. These c.v. obtained through D.A.
correspond to invariant solution (or at least to a particular solution) and
therefore they are induced by one symmetry. We compare our tactic with the
Lie one.

\hrulefill

\begin{example}
Solve the homogeneous equation
\begin{equation}
\left( x^{2}+y^{2}\right) dx=2xydy.  \label{ej1_ec1}
\end{equation}
\end{example}

\begin{proof}[Solution]
In order to solve it we will use three different ways, the traditional, the
dimensional and the Lie method.

\textbf{Traditional method. }Making the c.v. $u=y/x$ we have:
\begin{equation}
u^{\prime }=\frac{1-u^{2}}{2ux}\Longrightarrow \frac{2udu}{1-u^{2}}=\frac{dx%
}{x}\Longrightarrow \ln (1-u^{2})=\ln x\Longrightarrow y^{2}=x^{2}+x,
\end{equation}

\textbf{Dimensional Analysis. }We go next to consider eq. (\ref{ej1_ec1}),
written as follows
\begin{equation}
y^{\prime }=\frac{\left( a^{2}x^{2}+y^{2}\right) }{2xy},  \label{ej1_ec2}
\end{equation}%
where the dimensional constant $a,$ makes the ode verify the dimensional
principle of homogeneity (d.p.h.) if
\begin{equation}
\left[ a\right] =\left[ \frac{y}{x}\right] =X^{-1}Y,
\end{equation}%
where $\left[ \cdot \right] $ stands for the dimensional equation of the
quantity $\cdot .$

Applying the Pi theorem we obtain the dimensionless variables that help us
to simplify the original ode. Therefore taking into account the following
dimensional matrix we take
\begin{equation}
\begin{array}{r|rrr}
& y & x & a \\ \hline
x & 0 & 1 & -1 \\
y & 1 & 0 & 1%
\end{array}
\Longrightarrow \pi _{1}=\frac{ax}{y},\qquad \Longrightarrow y=ax,
\label{ej1_m1}
\end{equation}
as we will see later this solution is a particular solution of this ode. We
would like to emphasize that as we have only needed one constant then the
equation is scale-invariant in such a way that the generator of this group
is $X=x\partial _{x}+y\partial _{y}$ as it is observed from eq. (\ref{ej1_m1}%
). This fact will be probed studying this equation under the Lie group
tactic, see below.

In this way the new variables are $\left( t,u(t)\right) :$
\begin{equation}
\left( t=x,\qquad u(t)=a\frac{x}{y}\right) ,\hspace{5mm}\Longrightarrow
\left( x=t,\qquad y=a\frac{t}{u(t)}\right) ,
\end{equation}%
this change of variables brings us to rewrite eq. (\ref{ej1_ec2}) as
follows:
\begin{equation}
\frac{u^{\prime }}{u\left( 1-u^{2}\right) }=\frac{1}{2t}\qquad
\Longrightarrow \qquad \frac{u}{\sqrt{1-u^{2}}}=\frac{1}{2}\ln t,
\end{equation}%
and hence
\begin{equation}
y^{2}=a^{2}x^{2}+C_{1}x.
\end{equation}%
As we can see in this trivial example, the D.A. induces a c.v. which helps
us to obtain an ode simpler than the original one.

We can also think in the following way
\begin{equation}
ay^{\prime }=b\frac{x}{2y}+a\frac{y}{2x},  \label{claudia}
\end{equation}%
where
\begin{equation}
\left[ a\right] =x,\qquad \left[ b\right] =y^{2}x^{-1},
\end{equation}%
and hence
\begin{equation}
\left( t=\frac{x}{a},\qquad u(t)=\frac{y^{2}}{bx}\right) \Longrightarrow
\left( x=at,\qquad y=\sqrt{abtu(t)}\right) ,  \label{cv1}
\end{equation}%
therefore eq. (\ref{claudia}) yields
\begin{equation}
u^{\prime }=1\Longrightarrow u=t+C_{1},
\end{equation}%
in this way we obtain the solution
\begin{equation}
y^{2}=\frac{b}{a}x^{2}+C_{1}x,  \label{sol1}
\end{equation}%
once we have obtained the solution, the constants $a,b$ are makig equal to $%
1,$ i.e. $a=b=1.$

In the same way we may consider the following change of variables
\begin{equation}
\left( t=\frac{x}{a},\qquad u(t)=\frac{y}{\sqrt{bx}}\right) \Longrightarrow
\left( x=at,\qquad y=u(t)\sqrt{abt}\right) ,  \label{cv2}
\end{equation}
which brings us to the following ode
\begin{equation}
2u^{\prime }u=1\Longrightarrow u^{2}-t+C_{1}=0,
\end{equation}
and therefore we obtain again the solution (\ref{sol1}). As we will see
below all these c.v. are generated by their corresponding (respective)
symmetry.

As we can see, this last change of variables is better than the first and
the tactic is the same: to introduce dimensional constants that make the
equation dimensional homogeneous.

\textbf{Lie Method. }In order to find the symmetry generator of a ode
\begin{equation}
y^{\prime }=f\left( x,y\right) ,
\end{equation}%
we need to solve the following pde
\begin{equation}
\eta _{x}+({\eta }_{y}-{\xi }_{x})f-{\xi }_{y}f^{2}-\xi (x,y)f_{x}-\eta
(x,y)f_{y}=0,
\end{equation}%
where
\begin{equation}
f_{x}=\frac{df}{dx},\quad f_{y}=\frac{df}{dy}.
\end{equation}

In this case we have to solve
\begin{equation}
\eta _{x}+\frac{1}{2}\frac{({\eta _{y}}-{\xi }_{x})(x^{2}+y^{2})}{x\,y}-%
\frac{1}{4}\frac{{\xi _{y}}(x^{2}+y^{2})^{2}}{x^{2}\,y^{2}}-\xi (x,y)(\frac{1%
}{y}-\frac{x^{2}+y^{2}}{2\,x^{2}y})-\eta (x,y)(\frac{1}{x}-\frac{x^{2}+y^{2}%
}{2\,x\,y^{2}})=0,
\end{equation}%
we find that $\left( X=\xi \partial _{x}+\eta \partial _{y}\right) $:%
\begin{eqnarray}
X_{1} &=&\frac{x}{2\,y}\partial _{y},\qquad X_{2}=\left( -\frac{x^{2}}{2\,y}+%
\frac{y}{2}\right) \partial _{y},\qquad X_{3}=x\partial _{x}+y\partial _{y},
\notag \\
X_{4} &=&\partial _{x}+\frac{y}{2\,x}\partial _{y},\qquad X_{5}=\left(
x^{2}+y^{2}\right) \partial _{x}+2xy\partial _{y},
\end{eqnarray}%
observing that $X_{5}$ is a trivial symmetry.

Each symmetry induces a change of variable (canonical variables) which are
obtained through the following formula
\begin{equation}
Xt=0,\qquad Xu=1.
\end{equation}%
In this way the change of variables that induces the field $X_{4}=\partial
_{x}+\frac{y}{2\,x}\partial _{y}$ is:
\begin{equation}
\left( t=\frac{y}{\sqrt{x}},u(t)=x\right) \Longrightarrow \left( x=u(t),y=t%
\sqrt{u(t)}\right) ,
\end{equation}%
in such a way that eq. (\ref{ej1_ec1}) yields:
\begin{equation}
u^{\prime }=2t\Longrightarrow u(t)=t^{2}+C_{1},
\end{equation}%
and therefore
\begin{equation}
x=\frac{y^{2}}{x}+C_{1},
\end{equation}%
which is the same solution. The c.v. induced by this symmetry is similar to
the obtained one in (\ref{cv2}).

The c.v. that induces the field $X_{1}=\frac{x}{2\,y}\partial _{y}$ is the
following one
\begin{equation}
\left( t=x,\quad u(t)=\frac{y^{2}}{x}\right) \Longrightarrow \left(
x=t,\quad y=\sqrt{u(t)t}\right) ,
\end{equation}%
therefore eq.(\ref{ej1_ec1}) yields
\begin{equation}
u^{\prime }=b^{2},
\end{equation}%
this c.v. is the same than the obtained one in (\ref{cv1}).

For example the c.v. that induces the field $X_{3}=x\partial _{x}+y\partial
_{y}$ is
\begin{equation}
\left( t=\frac{y}{x},u(t)=\ln x\right) \Longrightarrow \left(
x=e^{u(t)},y=te^{u(t)}\right) ,
\end{equation}%
which brings us to the following ode :
\begin{equation}
u^{\prime }=-\frac{2t}{t^{2}-1}\Longrightarrow u(t)=-\ln \left( -1+t\right)
-\ln \left( 1+t\right) +C_{1},
\end{equation}%
now writing the solution in the original variables we take
\begin{equation}
\ln x=C_{1}+\ln \left( \frac{x}{\left( y-x\right) \left( y+x\right) }\right)
.
\end{equation}

Now if we consider the invariants that induce each symmetry%
\begin{equation}
\frac{dx}{\xi }=\frac{dy}{\eta },\qquad \longmapsto \qquad y^{\prime }:=%
\dfrac{dy}{dx}=\dfrac{\eta }{\xi },
\end{equation}%
we take that:%
\begin{equation}
X_{1}\longmapsto I_{1}=x,\qquad X_{2}\longmapsto I_{2}=x,\qquad
X_{3}\longmapsto I_{3}=\dfrac{y}{x},\qquad X_{4}\longmapsto I_{4}=\dfrac{y}{%
\sqrt{x}}.
\end{equation}

For example the symmetry $X_{3}=x\partial _{x}+y\partial _{y}$ generates the
following invariant:
\begin{equation}
\frac{dx}{\xi }=\frac{dy}{\eta }\text{\qquad }\Longrightarrow \qquad \frac{dx%
}{x}=\frac{dy}{y}\Longrightarrow \ln x=\ln y\Longrightarrow I_{3}=\dfrac{y}{x%
}\longmapsto y=ax,
\end{equation}%
this would be the solution that suggests us precisely the direct use of the
Pi theorem (if the ode is scale invariant, as in this case, the solution
obtained applying by the Pi theorem coincides with the invariant solution
that generates the scale symmetry, in this case $X_{3}$). We see that we
only obtain a particular solution, but that this is invariant, in fact if we
think about the ode as a dynamical system we see that the fixed point of
such equation would be precisely the solution $y=\pm x.$
\end{proof}

\hrulefill

\begin{example}
Solve the linear ode
\begin{equation}
\left( 1-x^{2}\right) y^{\prime }+xy=1.  \label{ej3_ec1}
\end{equation}
\end{example}

\begin{proof}[Solution]
\textbf{Solution through D.A. }Our first step will be to introduce
dimensional constants in such a way that eq. (\ref{ej3_ec1}) verifies the
principle of dimensional homogeneity. In this way we rewrite the equation as
follows
\begin{equation}
\left( A-x^{2}\right) y^{\prime }+xy=B,  \label{ej3_ec2}
\end{equation}%
where
\begin{equation}
\left[ A\right] =X^{2},\qquad \left[ B\right] =XY,  \label{ej3_m1}
\end{equation}%
i.e.
\begin{equation}
\pi _{1}=\frac{x^{2}}{A}\qquad \pi _{2}=\frac{xy}{B},
\end{equation}%
where
\begin{equation}
y=\dfrac{1}{Bx}\varphi \left( \frac{x^{2}}{A}\right) ,
\end{equation}%
being $\varphi $ a unknown function. It is possible to make the following
assumption
\begin{equation}
y=\dfrac{1}{Bx}\left( \frac{x^{2}}{A}\right) ^{n},\qquad \qquad n\in \mathbb{%
R}.
\end{equation}

Since $y=x^{-1}$ is not a particular solution of (\ref{ej3_ec1}) we need to
combine the $\pi -monomias$ in order to find a particular solution, the
simplest way is as follows:
\begin{equation}
\left( t=\frac{x^{2}}{A},\qquad u(t)=\frac{A}{B}\frac{y}{x}\right)
\Longrightarrow \left( x=\sqrt{At},\qquad y=\frac{B}{A}u(t)\sqrt{At}\right)
\end{equation}%
observing that
\begin{equation}
u(t)=\pi _{2}\cdot \pi _{1}^{-1},
\end{equation}%
where $y=cx$, derived from $\pi _{2}\cdot \pi _{1}^{-1}$ is a particular
solution of (\ref{ej3_ec1}).

\begin{remark}[Recipe]
When we have two $\pi -monomia,$ we have to check if they induce any
particular (or invariant) solution. If they are not particular solutions
then we must combine them in order to obtain such solution, this solution
usually is obtained by combining them in a very simple way.
\end{remark}

In this way our ode is written now as follows:
\begin{equation}
\frac{2u^{\prime }}{u-1}=\frac{1}{t\left( t-1\right) }\Longrightarrow
u=1+C_{1}\frac{\sqrt{t+1}}{t},
\end{equation}%
and in the original variables the solution of eq. (\ref{ej3_ec1}) yields:
\begin{equation}
y=\frac{A}{B}\left( x\pm C_{1}\sqrt{A-x^{2}}\right) ,
\end{equation}%
where $C_{1}$ is an integration constant. Now making $A=B=1$, we have the
ordinary solution to eq. (\ref{ej3_ec1}).

\textbf{Lie method. }Following the standard procedure we have to solve the
pde:
\begin{equation}
\eta _{x}+\frac{({\eta _{y}}-{\xi _{x}})\,(1-x\,y)}{1-x^{2}}-\frac{{\xi _{y}}%
\,(1-x\,y)^{2}}{(1-x^{2})^{2}}-\xi (x,\,y)\,(-\frac{y}{1-x^{2}}+\frac{%
2\,(1-x\,y)\,x}{(1-x^{2})^{2}})+\frac{\eta (x,\,y)\,x}{1-x^{2}}=0,
\end{equation}%
which solutions are:%
\begin{equation}
X_{1}=\sqrt{-1+x^{2}}\partial _{y},\qquad X_{2}=\left( -y+x\right) \partial
_{y},\qquad X_{3}=\left( \frac{(2xy-y^{2}-1)(-y+x)}{(x-1)(x+1)}\right)
\partial _{y},
\end{equation}%
and their corresponding invariants are:%
\begin{equation}
X_{1}\longmapsto I_{1}=x,\qquad X_{2}\longmapsto I_{2}=x,\qquad
X_{3}\longmapsto I_{3}=x.
\end{equation}

For example the symmetry $X_{1}$ generates the following c.v.:
\begin{equation}
\left( t=x,\quad u(t)=\frac{y}{\sqrt{-1+x^{2}}}\right) \Longrightarrow
\left( x=t,\quad y=u(t)\sqrt{-1+t^{2}}\right) ,
\end{equation}%
therefore eq. (\ref{ej3_ec1}) is written as:
\begin{equation}
u^{\prime }=-\frac{\sqrt{-1+t^{2}}}{1-2t^{2}+t^{4}},
\end{equation}%
which solution is:
\begin{equation}
u(t)=-\frac{\left( -1+t^{2}\right) ^{3/2}}{4\left( t+1\right) ^{2}}+\frac{%
\left( -1+t^{2}\right) ^{3/2}}{4\left( t-1\right) ^{2}}+C_{1},
\end{equation}%
and hence
\begin{equation}
\frac{y}{\sqrt{x^{2}-1}}=\frac{C_{1}(x^{2}-1)+x\sqrt{x^{2}-1}}{x^{2}-1}%
\Longrightarrow y=x\pm c\sqrt{x-1}\sqrt{x+1}.
\end{equation}

Now, if for example we consider the symmetry $X_{2}$ then we have:
\begin{equation}
\left( t=x,\quad u(t)=-\ln (-y+x)\right) \Longrightarrow \left( x=t,\quad y=%
\frac{te^{u(t)}-1}{e^{u(t)}}\right) ,
\end{equation}%
therefore:
\begin{equation}
u^{\prime }=-\frac{t}{-1+t^{2}},
\end{equation}%
finding in this way that its solution is:
\begin{equation}
u(t)=-\frac{1}{2}\ln (t-1)-\frac{1}{2}\ln (t+1)+C_{1},
\end{equation}%
hence in the original variables the solution yields:
\begin{equation}
-\ln (-y+x)=-\frac{1}{2}\ln (x-1)-\frac{1}{2}\ln (x+1)+C_{1},
\end{equation}%
which after a simple simplification it yields:
\begin{equation}
y=x\pm c\sqrt{x-1}\sqrt{x+1},
\end{equation}%
as we already know.
\end{proof}

\hrulefill

With these two simple examples we have tried to show how the D.A. works in
order to introduce c.v. which brings us to obtain simpler odes than the
original one. As we have emphasized in the recipe, the trick is to look for
a particular solution. Sometimes this particular solution will be
furthermore invariant solution (induced by a concrete symmetry). If this is
the case, then our ode will be reduced to an ode with separating variables
but if this solution is only a particular then, as we will see below, we
have not any guarantee of reducing our ode to an ode with separate
variables, nevertheless we will obtain a simpler ode than the original one.
In the next section we will study some examples.

\section{Examples}

In this section we will apply our pedestrian method to different odes,
beginning with a Bernoulli ode. We go next to employ the method to solve
Abel odes as well as Riccati odes since as anyone knows these kind of odes
are truly very difficult.

\hrulefill

\begin{example}
Solve the following Bernoulli ode.
\begin{equation}
y^{\prime }=-\frac{y}{x+1}-\frac{\left( x+1\right) y^{2}}{2}.
\label{Bernoulli}
\end{equation}
\end{example}

\begin{remark}
Historically it was Bernoulli the first person who introduced c.v. in order
to solve odes (now bearing his name). He managed to reduce this equation to
a simpler linear equation.
\end{remark}

\begin{proof}[Solution]
It is observed that with
\begin{equation}
\left[ a\right] =X,\qquad \left[ b\right] =X^{-2}Y^{-1},
\end{equation}%
eq. (\ref{Bernoulli}) yields d.h. i.e.
\begin{equation}
y^{\prime }=-\frac{y}{x+a}-\frac{b\left( x+a\right) y^{2}}{2},
\end{equation}%
in this way it is obtained the following variables
\begin{equation}
\left( t=\frac{x}{a},\qquad u(t)=\frac{1}{abxy}\right) \Longrightarrow
\left( x=at,\qquad y=\frac{1}{a^{2}btu}\right) ,
\end{equation}%
with this change of variables eq. (\ref{Bernoulli}) yields
\begin{equation}
2u+2t\left( 1+t\right) u^{\prime }=\left( t+1\right) ^{2},
\end{equation}%
which is linear and its solution is:
\begin{equation}
u=\left( \frac{t}{2}+C_{1}\right) \left( 1+\frac{1}{t}\right) ,
\end{equation}%
and therefore in the original variables the solution yields
\begin{equation}
y=\frac{2}{\left( bx+C_{1}\right) \left( x+a\right) },
\end{equation}%
in this case D.A. does not bring us to obtain a good change of variables but
helps us to obtain a simpler ode than the original one. We can try now with
the following c.v.
\begin{equation}
\left( t=x,\qquad u(t)=\frac{1}{x^{2}y}\right) \Longrightarrow \left(
x=t,\qquad y=\frac{1}{t^{2}u}\right) ,
\end{equation}%
which brings us to obtain a linear ode
\begin{equation}
t(t+1)\left( 2u^{\prime }t+4u\right) =2t^{2}u+t^{2}+2t+1,
\end{equation}%
but with this tactic we are not able to obtain a simpler ode.

As we can see D.A. induces us to the c.v. $\frac{1}{abxy}$ but $y=\dfrac{1}{x%
}$ or $y=\dfrac{1}{x^{2}},$ are not particular solutions of (\ref{Bernoulli}%
). Nevertheless solutions as $y=\dfrac{1}{x+1}$ or $y=\dfrac{1}{\left(
x+1\right) ^{2}}$ are particular and invariant solutions for this reason
these c.v. induce us to obtain an ode in separate variables, but
unfortunately D.A. is unable to construct such c.v.. For this reason we can
only reduced our ode to a linear ode as in the theoretical case i.e.
employing the theoretical method purposed by Bernoulli.

If we study this ode under the Lie group method, it is observed that eq. (%
\ref{Bernoulli}) admits the following symmetries obtained from:%
\begin{equation*}
\eta _{x}+(\eta _{y}-\xi _{x})\left( -\frac{y}{x+1}-\frac{\left( x+1\right)
y^{2}}{2}\right) -\xi _{y}\left( -\frac{y}{x+1}-\frac{\left( x+1\right) y^{2}%
}{2}\right) ^{2}-
\end{equation*}%
\begin{equation}
-\xi \left( \dfrac{y}{\left( x+1\right) ^{2}}-\dfrac{1}{2}y^{2}\right)
\,-\eta \left( -y-xy-\dfrac{1}{x+1}\right) =0
\end{equation}%
\begin{equation}
X_{1}=\frac{y^{2}\left( x+1\right) }{2}\partial _{y},\qquad X_{2}=\left(
x+1\right) \partial _{x}-2y\partial _{y},\qquad X_{3}=\partial _{x}-\dfrac{y%
}{\left( x+1\right) }\partial _{y},\qquad X_{4}=x\partial _{x}-\dfrac{\left(
1+2x\right) y}{\left( x+1\right) }\partial _{y},
\end{equation}%
and which respective invariants are:%
\begin{equation*}
I_{1}=x,\qquad I_{2}=y\left( x+1\right) ^{2},\qquad I_{3}=y\left( x+1\right)
,\qquad I_{4}=yx\left( x+1\right) ,
\end{equation*}%
that induces the following canonical variables. For example from $X_{1}:$
\begin{equation}
\left( t=x,\qquad u(t)=-\frac{2}{y\left( x+a\right) }\right) \Longrightarrow
\left( x=t,\qquad y=-\frac{2}{u\left( t+a\right) }\right) ,
\end{equation}%
in such a way that eq. (\ref{Bernoulli}) yields
\begin{equation}
u^{\prime }=-b\Longrightarrow y=bt+C_{1},
\end{equation}%
and in the original variables i.e. $\left( x,y\right) $ this solution yields
\begin{equation}
y=\frac{2}{\left( bx+C_{1}\right) \left( x+a\right) },
\end{equation}%
as we already know.
\end{proof}

\hrulefill

\begin{example}
Solve the following Riccati ode
\begin{equation}
y^{\prime }=xy^{2}-\frac{2y}{x}-\frac{1}{x^{3}}.  \label{Ricc1}
\end{equation}
\end{example}

\begin{proof}[Solution]
It is observed that introducing the following dimensional constants eq. (\ref%
{Ricc1}) verifies the principle of dimensional homogeneity (p.d.h.),
\begin{equation}
y^{\prime }=\frac{xy^{2}}{a}-\frac{2y}{x}-\frac{a}{x^{3}},
\end{equation}%
where $\left[ a\right] =X^{2}Y.$ As we can see if we need only one constant
then our equation is scale invariant and therefore this constant indicates
us that the generator of this symmetry is $X=-2x\partial _{x}+y\partial _{y}$
as we will see below. Therefore we have the following variables
\begin{equation}
\left( t=x,\qquad u(t)=\frac{a}{yx^{2}}\right) \Longrightarrow \left(
x=t,\qquad y=\frac{a}{tu^{2}}\right) ,
\end{equation}%
with this new variables eq. (\ref{Ricc1}) yields
\begin{equation}
\frac{u^{\prime }}{u^{2}-1}=\frac{1}{t}\Longrightarrow \ln t+\mathrm{arctanh}%
(u)+C_{1}=0,
\end{equation}%
and in the original variables this solutions is written as:
\begin{equation}
\ln x+\mathrm{arctanh}(\frac{a}{yx^{2}})+C_{1}=0.
\end{equation}

We would like to emphasize that in this case, as we have only one
dimensional constant, we can find the particular solution $y=ax^{-2}$
applying the Pi theorem, and that such solution (invariant particular
solution) is induced by the scale symmetry.

Eq. (\ref{Ricc1}) equation has the following symmetries:
\begin{equation}
\eta _{x}+(\eta _{y}-\xi _{x})(x\,y^{2}-\frac{2\,y}{x}-\frac{1}{x^{3}})-\xi
_{y}(xy^{2}-\frac{2\,y}{x}-\frac{1}{x^{3}})^{2}-\xi \,(y^{2}+\frac{2\,y}{%
x^{2}}+\frac{3}{x^{4}})-\eta \left( 2xy-\frac{2}{x}\right) =0
\end{equation}%
and hence%
\begin{eqnarray}
X_{1} &=&\dfrac{1}{x}\partial _{x}-\frac{2}{x^{4}}\partial _{y},\qquad
X_{2}=x^{4}\,(y+\frac{1}{x^{2}})^{2}\partial _{y},\qquad ,X_{3}=x\partial
_{x}-2y\,\partial _{y}  \notag \\
X_{4} &=&\left( -\frac{1}{x^{2}}+y^{2}\,x^{2}\right) \partial _{y},\qquad
X_{5}=x^{3}\partial _{x}-\left( 2+4y\,x^{2}\right) \partial _{y}.
\end{eqnarray}%
and which corresponding invariants are:%
\begin{equation}
I_{1}=\dfrac{yx^{2}-1}{x^{2}},\qquad I_{2}=I_{4}=x,\qquad
I_{3}=yx^{2},\qquad I_{5}=yx^{2}\left( x^{2}+1\right) .
\end{equation}

The method of canonical variables brings us to obtain the following ode, for
example, for the symmetry $X_{5}$, we have
\begin{equation}
\left( t=x^{2}\,(y\,x^{2}+1),\qquad u(t)=-\frac{1}{2\,x^{2}}\right) ,
\end{equation}%
\begin{equation}
u_{t}=\frac{1}{t^{2}}\Longrightarrow u(t)=-\frac{1}{t}+C_{1},
\end{equation}%
and hence
\begin{equation}
-\frac{1}{2\,x^{2}}=-\frac{1}{x^{2}\,(y\,x^{2}+1)}+C_{1},
\end{equation}%
as we already know.

The transformation induced by $X_{3}$ (scaling symmetry) is the following
one:
\begin{equation}
\left( t=y\,x^{2},\,\qquad u(t)=\ln (x)\right) \Longrightarrow \left(
x=e^{u(t)},\,\qquad y=\frac{t}{(e^{u(t)})^{2}}\right) ,
\end{equation}%
\begin{equation}
u_{t}=\frac{1}{t^{2}-1}\Longrightarrow u(t)=-\mathrm{arctanh}(t)+C_{1},
\end{equation}%
therefore
\begin{equation}
\ln (x)=-\mathrm{arctanh}(y\,x^{2})+C_{1},
\end{equation}%
which looks a little different than the other solution.
\end{proof}

\hrulefill

\begin{example}
Solve the following Riccati ode:
\begin{equation}
y^{\prime }=\frac{y+1}{x}+\frac{y^{2}}{x^{3}}.  \label{Ricc2}
\end{equation}
\end{example}

\begin{proof}[Solution]
The ode verifies the p.d.h. if we introduce the following dimensional
constant:
\begin{equation}
y^{\prime }=\frac{y+a}{x}+b\frac{y^{2}}{x^{3}},
\end{equation}%
where $\left[ a\right] =Y,\left[ b\right] =X^{2}Y^{-1}.$ In this way it is
obtained the c.v.
\begin{equation}
\left( t=\frac{ax^{2}}{by^{2}},\qquad u(t)=\frac{y}{a}\right)
\Longrightarrow \left( x=\sqrt{abtu^{2}},\qquad y=au\right) ,
\end{equation}%
and therefore eq. (\ref{Ricc2}) yields
\begin{equation}
\frac{u}{u^{\prime }}+t=\frac{ut}{\left( ut+t+1\right) },
\end{equation}%
which is a Bernoulli ode and its solution is:
\begin{equation}
\frac{1}{u}-\sqrt{t}\left( \arctan \sqrt{t}+C_{1}\right) =0,
\end{equation}%
hence in the original variables the solution is
\begin{equation}
\frac{a}{y}-\sqrt{\frac{ax^{2}}{by^{2}}}\left( \arctan \sqrt{\frac{ax^{2}}{%
by^{2}}}+C_{1}\right) =0.
\end{equation}%
It is observed that we have the particular solution $y=Cx,$ obtained from
the relationship $\frac{ax^{2}}{by^{2}}.$

Since this c.v. is not very good for our purposes we may now proceed as
follows:
\begin{equation}
axy^{\prime }=ay+1+b^{2}\frac{y^{2}}{x^{2}}
\end{equation}%
where
\begin{equation}
\left[ a\right] =y^{-1},\qquad \left[ b\right] =y^{-1}x
\end{equation}%
in such a way that these constants induce the following change of variables:
\begin{equation}
\left( t=\frac{a}{b}x,\qquad u(t)=\frac{x}{by}\right) \Longrightarrow \left(
x=\frac{b}{a}x,\qquad y=\frac{t}{au(t)}\right)
\end{equation}%
which brings us to obtain
\begin{equation}
-\frac{u^{\prime }}{u^{2}+1}=\frac{1}{t^{2}}\Longrightarrow -\frac{1}{t}%
+\arctan (u)+C_{1}=0
\end{equation}%
and hence
\begin{equation}
-\frac{b}{ax}+\arctan (\frac{x}{by})+C_{1}=0
\end{equation}%
which is a better approximation. In the same way we can obtain a quadrature
simply considering $t=\dfrac{b}{ax}$ and $u(t)=\frac{x}{by}.$ in this way it
is obtained $u^{\prime }=u^{2}+1.$

The Lie method bring us to solve the following pde
\begin{equation}
\eta _{x}+(\eta _{y}-\xi _{x})(\frac{y+1}{x}+\frac{y^{2}}{x^{3}})-\xi _{y}(%
\frac{y+1}{x}+\frac{y^{2}}{x^{3}})^{2}-\xi (-\frac{y+1}{x^{2}}-\frac{3\,y^{2}%
}{x^{4}})-\eta (\frac{1}{x}+\frac{2\,y}{x^{3}})=0,
\end{equation}%
and which solutions are:%
\begin{eqnarray}
X_{1} &=&\left( x+\frac{y^{2}}{x}\right) \partial _{y},\qquad X_{2}=\left(
\frac{(y\mathrm{cos}({\frac{1}{x}})+x\mathrm{sin}({\frac{1}{x}}))^{2}}{x}%
\right) \partial _{y}  \notag \\
X_{3} &=&\left( -\frac{\mathrm{sin}({\frac{1}{x}})^{2}\,(y\,\mathrm{sin}({%
\frac{1}{x}})-\mathrm{cos}({\frac{1}{x}})\,x)^{2}}{x\,(-1+\mathrm{cos}({%
\frac{1}{x}})^{2})}\right) \partial _{y},\qquad X_{4}=x^{2}\partial
_{x}+xy\partial _{y}
\end{eqnarray}%
and which invariants are:
\begin{equation}
I_{1}=I_{2}=I_{3}=x,\qquad I_{4}=\dfrac{y}{x}.
\end{equation}

The symmetry $X_{1}=\left( x+\frac{y^{2}}{x}\right) \partial _{y}$ generates
the following c.v.:
\begin{equation}
\left( t=x,\qquad u(t)=\arctan \left( \frac{y}{x}\right) \right)
\Longrightarrow \left( y=\tan (u(t))\,t,\,\qquad x=t\right) ,
\end{equation}%
\begin{equation}
u_{t}=\frac{1}{t^{2}}\Longrightarrow u(t)=-\frac{1}{t}+C_{1},
\end{equation}%
and hence the solution is
\begin{equation}
\arctan \left( \frac{y}{x}\right) =-\frac{1}{x}+C_{1},
\end{equation}%
as we already know.

We would like to emphasize that this ode is not scale invariant and
nevertheless we have been able to obtain a c.v. that reduces the ode to a
quadrature.
\end{proof}

\hrulefill

\begin{example}
Solve the following Abel ode
\begin{equation}
y^{\prime }=\frac{1-y^{2}}{x\,y}+1  \label{Abel1}
\end{equation}
\end{example}

\begin{proof}[Solution]
Introducing the following dimensional constants, we make that eq. (\ref%
{Abel1}) verifies the p.d.h.%
\begin{equation}
\left( xy\right) y^{\prime }=-y^{2}+bxy+a,
\end{equation}%
where $\left[ b\right] =yx^{-1},\left[ a\right] =y^{2}$. Therefore the c.v.
that suggests the D.A. is the following one:
\begin{equation}
\left( t=\frac{bxy}{a},\qquad u(t)=\frac{y^{2}}{a}\right) \Longrightarrow
\left( x=\frac{at}{b\sqrt{ua}},\qquad y=\sqrt{ua}\right) ,
\end{equation}%
hence eq. (\ref{Abel1}) yields
\begin{equation}
u^{\prime }\left( 2+t-u\right) t=2u\left( 1+t-u\right) ,
\end{equation}%
and its solution is:
\begin{equation}
u=\frac{t^{2}}{-2t+2\ln (1+t)-C_{1}},
\end{equation}%
in the original variables it yields:
\begin{equation}
-\frac{b^{2}x^{2}}{2a}=-\frac{bxy}{a}+\ln (1+\frac{bxy}{a})+C_{1}.
\end{equation}

Once again we emphasize that the particular solution, $y=Cx^{-1},$ (in this
case invariant solution , see below) has been obtained from the relationship
$\frac{bxy}{a}$.

Alternatively we can try the following c.v. Writing the original Abel ode,
eq.(\ref{Abel1}) in the following form:
\begin{equation}
axyy^{\prime }=-\frac{y^{2}}{a}+xy+b,  \label{paula}
\end{equation}%
where $\left[ a\right] =XY^{-1},\left[ b\right] =XY$, therefore we find the
next c.v.
\begin{equation}
\left( t=\frac{xy}{b},\quad u(t)=\frac{x^{2}}{ab}\right) \Longrightarrow
\left( x=\sqrt{abu},\quad y=\frac{bt}{\sqrt{abu}}\right) ,
\end{equation}%
that brings us to rewrite eq. (\ref{paula}) as follows:
\begin{equation}
u^{\prime }=\frac{2t}{1+t}\Longrightarrow u=2t-2\ln (1+t)+C_{1},
\end{equation}%
and hence in the original variables $\left( x,y\right) $ we get:
\begin{equation}
\frac{x^{2}}{ab}=2\frac{xy}{b}-2\ln (1+\frac{xy}{b})+C_{1},
\end{equation}%
as we already know. We can check that the following c.v. also works well
\begin{equation}
\left( t=\frac{x^{2}}{ab},\quad u(t)=\frac{xy}{b}\right) \Longrightarrow
\left( x=\sqrt{abt},\quad y=\frac{bu}{\sqrt{abt}}\right) ,
\end{equation}%
since in these variables eq. (\ref{paula}) is written as
\begin{equation}
2u^{\prime }u=1+u\Longrightarrow t-2u+2\ln (1+u)+C_{1}=0,
\end{equation}%
and therefore
\begin{equation}
\frac{x^{2}}{ab}-2\frac{xy}{b}+2\ln (1+\frac{xy}{b})+C_{1}=0,
\end{equation}%
obtaining the solution.

Applying the Lie method we need to solve the following pde:
\begin{equation}
\eta _{x}+(\eta _{y}-\xi _{x})(\frac{a-y^{2}}{x\,y}+1)-\xi _{y}\,(\frac{%
a-y^{2}}{x\,y}+1)^{2}+\frac{\xi \,(a-y^{2})}{x^{2}\,y}+\eta (\frac{2}{x}+%
\frac{a-y^{2}}{xy^{2}})=0,
\end{equation}%
which solution is:
\begin{equation}
X_{1}=-\frac{1}{x}\partial _{x}+\frac{y}{x^{2}}\partial _{y}\qquad
\Longrightarrow \qquad I_{1}=xy.
\end{equation}

This symmetry induces the following c.v.:
\begin{equation}
\left( u(t)=-\frac{x^{2}}{2},\qquad \,t=x\,y\right) ,
\end{equation}%
and therefore in this variables eq. (\ref{paula}) is written as follows:
\begin{equation}
u^{\prime }=-\frac{t}{a+t}\Longrightarrow u(t)=-t+a\,\ln (a+t)+C_{1},
\end{equation}%
hence in the original variables we get:
\begin{equation}
-\frac{x^{2}}{2}=-x\,y+a\,\ln (a+x\,y)+C_{1},
\end{equation}

This is another example of an ode that it is not scale invariant, and
nevertheless, we have been able to reduce it to a quadrature.
\end{proof}

\hrulefill

\begin{example}
Solve the following Abel ode.
\begin{equation}
x\left( y+4\right) y^{\prime }=y^{2}+2y+2x.  \label{Abel2}
\end{equation}
\end{example}

\begin{proof}[Solution]
Following our pedestrian method, we begin by introducing dimensional
constants and rewriting the odes as follows
\begin{equation}
x\left( y+4a\right) y^{\prime }=y^{2}+2ay+2bx,
\end{equation}%
where $\left[ a\right] =y,$\ and \ $\left[ b\right] =x^{-1}y^{2},$ therefore
D.A. suggest us the following c.v. (note that $y=\sqrt{x},$ is a particular
solution of (\ref{Abel2}))
\begin{equation}
\left( t=\frac{y^{2}}{bx},\qquad u(t)=\frac{y}{a}\right) \Longrightarrow
\left( x=\frac{u^{2}a^{2}}{ct},\qquad y=ua\right) ,
\end{equation}%
in such a way that eq. (\ref{Abel2}) yields
\begin{equation}
t\left( t+4\right) u^{\prime }=t(u+2)+2u,
\end{equation}%
which is a linear ode and its solution is:
\begin{equation}
u=t+C_{1}\sqrt{t\left( t+4\right) },
\end{equation}%
hence
\begin{equation}
\frac{y}{a}=\frac{y^{2}}{bx}+C_{1}\sqrt{\frac{y^{2}}{bx}\left( \frac{y^{2}}{%
bx}+4\right) },
\end{equation}%
and simplifying, it is obtained the solution:
\begin{equation}
\frac{bx}{a}=y+C_{1}\sqrt{\left( y^{2}+4bx\right) }.
\end{equation}

Another particular solution could be found from the following relationship $%
\frac{y^{2}}{bx}\left( \frac{y}{a}\right) ^{-1}$ i.e. $y=Cx.$ But this
particular solution bring us to the following c.v.
\begin{equation}
\left( t=\frac{ay}{bx},\qquad u(t)=\frac{y}{a}\right) \Longrightarrow \left(
x=\frac{ua^{2}}{ct},\qquad y=ua\right) ,
\end{equation}%
which transforms eq. (\ref{Abel2}) into a Bernoulli ode,
\begin{equation}
t^{2}\left( u+4\right) u^{\prime }=\left( ut+2t+2\right) \left( u^{\prime
}t-u\right) ,
\end{equation}%
but this situation is not desirable since it is always more difficult to
solve a Bernoulli ode than a linear ode.

Applying the Lie method, following the standard procedure, we need to solve
the following pde
\begin{equation*}
\eta _{x}+\frac{({\eta _{y}}-{\xi _{x}})\,(y^{2}+2\,y+2\,x)}{x\,(y+4)}-\frac{%
{\xi _{y}}\,(y^{2}+2\,y+2\,x)^{2}}{x^{2}\,(y+4)^{2}}-
\end{equation*}%
\begin{equation}
-\xi \,(\frac{2}{x\,(y+4)}-\frac{y^{2}+2\,y+2\,x}{x^{2}\,(y+4)})-\eta (\frac{%
2\,y+2}{x\,(y+4)}-\frac{y^{2}+2\,y+2\,x}{x\,(y+4)^{2}})=0,
\end{equation}%
and which solutions are:%
\begin{equation}
X_{1}=\left( {\frac{(2\,y+4-x)\,(y-x)}{y+4}}\right) {\partial }_{y},\qquad
X_{2}=\left( \frac{(y^{2}+4\,x)\,(y-x)}{x\,(y+4)}\right) {\partial }%
_{y},\qquad X_{3}=\left( 4x+x^{2}\right) \partial _{x}+x\left( y+4\right)
\partial _{y}
\end{equation}%
where their correspond invariants are:%
\begin{equation}
I_{1}=I_{2}=x,\qquad I_{3}=\dfrac{y+4}{x+4}
\end{equation}

For example the c.v. that induces $X_{3}$ is:
\begin{equation}
\left( t=\frac{y+4}{4+x},\,u(t)=\frac{1}{4}\ln (x)-\frac{1}{4}\ln
(4+x)\right) ,
\end{equation}%
hence:
\begin{equation}
\left( y=-\frac{4\,(t-1+e^{(4\,u(t))})}{-1+e^{(4\,u(t))}},\,x=-\frac{%
4\,e^{(4\,u(t))}}{-1+e^{(4\,u(t))}}\right) \!,
\end{equation}%
finding that:
\begin{equation}
u^{\prime }=\frac{t}{2(2t^{2}-3t+1)},
\end{equation}%
which solution is:
\begin{equation}
u(t)=\frac{1}{2}\,\ln (t-1)-\frac{1}{4}\,\ln (2\,t-1)+C_{1},
\end{equation}%
therefore, in the original variables it yields:
\begin{equation}
\frac{1}{4}\,\ln (x)-\frac{1}{4}\,\ln (4+x)=\frac{1}{2}\,\ln (\frac{y+4}{4+x}%
-1)-\frac{1}{4}\,\ln (\frac{2\,(y+4)}{4+x}-1)+C_{1},
\end{equation}%
as we already know.
\end{proof}

\hrulefill

\begin{example}
Solve the Abel ode.
\begin{equation}
y^{\prime }=Cx^{3}y^{3}+Bxy^{2}-A\frac{y}{x},  \label{Abel3}
\end{equation}
\end{example}

\begin{proof}[Solution]
If we rewrite eq. (\ref{Abel3}) introducing the following dimensional
constants,
\begin{equation}
y^{\prime }=a^{2}Cx^{3}y^{3}+aBxy^{2}-A\frac{y}{x},
\end{equation}%
where $A,B,C\in \mathbb{R},$ and $\left[ a\right] =X^{-2}Y^{-1}$. As we can
see this ode is scale invariant since we have needed introduce only one
constant. D.A. suggests us the following c.v.
\begin{equation}
\left( t=x,\qquad u(t)=ax^{2}y\right) \Longrightarrow \left( x=t,\qquad y=%
\frac{u}{at^{2}}\right) ,
\end{equation}%
in such a way that eq. (\ref{Abel3}) is written in the following form:
\begin{equation}
tu^{\prime }=u\left( u^{2}+u+1\right) ,
\end{equation}%
and its solution is:
\begin{equation}
\ln t+\frac{1}{2}\ln \left( u^{2}+u+1\right) +\frac{\sqrt{3}}{3}\arctan
\left( \left( \frac{3}{2}u+\frac{1}{3}\right) \sqrt{3}\right) -\ln u+C_{1}=0,
\end{equation}%
therefore in the original variables it yields:
\begin{equation}
\ln x+\frac{1}{2}\ln \left( \left( ax^{2}y\right) ^{2}+ax^{2}y+1\right) +%
\frac{\sqrt{3}}{3}\arctan \left( \left( \frac{3}{2}\left( ax^{2}y\right) +%
\frac{1}{3}\right) \sqrt{3}\right) -\ln \left( ax^{2}y\right) +C_{1}=0.
\end{equation}

In second place, we study eq. (\ref{Abel3})
\begin{equation}
y^{\prime }=Cx^{3}y^{3}+Bxy^{2}-A\frac{y}{x},
\end{equation}%
with respect to the dimensional base $B=\left\{ T\right\} .$ This ode
verifies the principle of dimensional homogeneity with respect to this
dimensional base. Note that $\left[ y\right] =\left[ \frac{1}{H^{\prime }}%
\right] =T^{2},$ and $\left[ x\right] =\left[ H\right] =T^{-1}$ hence $\left[
y^{\prime }\right] =T^{3}.$ Therefore rewriting the equation in a
dimensionless way \ we find that $y\propto x^{-2}$

But if we study this equation with respect to the dimensional base $%
B=\left\{ X,Y\right\} ,$ we need to introduce new dimensional constants that
make the equation verify the principle of dimensional homogeneity
\begin{equation}
y^{\prime }=\alpha Cx^{3}y^{3}+\beta Bxy^{2}-A\frac{y}{x}
\end{equation}%
where $\left[ \alpha ^{1/2}\right] =\left[ \beta \right] =X^{-2}Y^{-1},$
hence
\begin{equation}
\begin{array}{r|rrr}
& y & \beta & x \\ \hline
X & 0 & -2 & 1 \\
Y & 1 & -1 & 0%
\end{array}%
\Longrightarrow y\propto \frac{\beta }{x^{2}},
\end{equation}

As we can see we have obtained the same solution than in the case of the
invariant solution. This is because the invariant solution that induces a
scaling symmetry is the same as the obtained one through the Pi theorem.

This ode admits the following symmetry (scale-invariant)
\begin{equation}
X=x\partial _{x}-2y\partial _{y},\qquad \Longrightarrow \qquad I=x^{2}y
\label{symabel1}
\end{equation}%
which is a scaling symmetry and it induces the following change of
variables,
\begin{equation}
r=x^{2}y,\quad s(r)=\ln (x),\quad \Longrightarrow \quad x=e^{s(r)},\quad y=%
\frac{r}{e^{2s(r)}},
\end{equation}%
which brings us to obtain the next ode in quadratures
\begin{equation}
s^{\prime }=\frac{1}{r\left( Cr^{2}+Br+2-A\right) },
\end{equation}%
and which solution is:
\begin{equation}
s(r)=-\frac{\ln r}{A-2}+\frac{1}{2}\frac{\ln \left( Cr^{2}+Br+2-A\right) }{%
A-2}-\frac{B\mathrm{arctanh}\left( \frac{2Cr+B}{\sqrt{B^{2}+4C(A-2)}}\right)
}{\left( A-2\right) \sqrt{B^{2}+4C(A-2)}}+C_{1},
\end{equation}%
and hence in the original variables $\left( x,y\right) $:
\begin{equation}
\ln x=-\frac{\ln \left( x^{2}y\right) }{A-2}+\frac{1}{2}\frac{\ln \left(
Cx^{4}y^{2}+Bx^{2}y+2-A\right) }{A-2}-\frac{B\mathrm{arctanh}\left( \frac{%
2Cx^{2}y+B}{\sqrt{B^{2}+4C(A-2)}}\right) }{\left( A-2\right) \sqrt{%
B^{2}+4C(A-2)}}+C_{1},
\end{equation}%
which is the most general solution for this ode.
\end{proof}

\hrulefill

\section{Pathological cases.}

In this section we will present two examples of odes that do not admit
symmetries (Lie point symmetries). Nevertheless in the first of them D.A.
helps us to obtain a simple c.v. that will bring us to obtain a simpler ode
(through a particular solution). In the second case we will show that
unfortunately sometimes one finds odes that at this time have no solution,
or at least we do not know how to solve them.

\hrulefill

\begin{example}
Solve the Abel ode
\begin{equation}
\left( x^{2}y+x^{5}-x\right) y^{\prime }=xy^{2}-\left( x^{4}+1\right) y.
\label{pato1}
\end{equation}
\end{example}

\begin{proof}[Solution]
Following our pedestrian method we beginn by introducing dimensional
constants%
\begin{equation}
ax^{2}yy^{\prime }+bx^{5}y^{\prime }-cxy^{\prime }=dxy^{2}-ex^{4}y+fy
\end{equation}%
where
\begin{equation}
\left[ a\right] =\left[ d\right] =x^{-2}y^{-1},\qquad \left[ b\right] =\left[
e\right] =x^{-5},\qquad \left[ c\right] =\left[ f\right] =x^{-1},
\end{equation}%
therefore $\left[ c^{5}\right] =\left[ f^{5}\right] =\left[ b\right] =\left[
e\right] ,$ having only two dimensional constants. Since $y=x^{-2}$ is not a
particular solution of eq.(\ref{pato1}) then we look for a combination
between the monomias finding in this way that
\begin{equation}
\pi _{1}=\dfrac{1}{cx},\qquad \pi _{2}=\dfrac{axy}{c}
\end{equation}%
where $y=x^{-1}$ is a particular solution of (\ref{pato1}). The c.v. that
induces the D.A. is the following one:
\begin{equation}
\left( t=xy,\quad u(t)=x\right) \Longrightarrow \left( x=u,\quad y=\dfrac{t}{%
u}\right) ,
\end{equation}%
hence using these new variables eq. (\ref{pato1}) is written now as:
\begin{equation}
2u^{\prime }t\left( t+1\right) +u\left( 1-t-u^{4}\right) =0,
\end{equation}%
which is a Bernoulli ode and its solution is%
\begin{equation}
u=\dfrac{\sqrt{t\sqrt{-2t-2\ln (t-1)+C_{1}}}}{\sqrt{-2t-2\ln (t-1)+C_{1}}},
\end{equation}%
undoing the c.v. we find that the solution to eq. (\ref{pato1}) is:%
\begin{equation}
x=\dfrac{\sqrt{xy\sqrt{-2xy-2\ln \left( xy-1\right) +C_{1}}}}{\sqrt{%
-2xy-2\ln \left( xy-1\right) +C_{1}}}.
\end{equation}%
$\allowbreak $

Applying the Lie method we see that the pde to solve is:
\begin{equation*}
\eta _{x}+({\eta _{y}}-{\xi _{x}})\frac{y\left( xy-x^{4}-1\right) }{x\left(
xy+x^{4}-1\right) }-{\xi }_{y}\frac{\,y^{2}\left( xy-x^{4}-1\right) ^{2}}{%
x\left( xy+x^{4}-1\right) ^{2}}-
\end{equation*}%
\begin{equation*}
-\xi \,\left( \frac{y\left( y-4x^{3}\right) }{x\left( xy+x^{4}-1\right) }-%
\frac{y\left( xy-x^{4}-1\right) }{x^{2}\left( xy+x^{4}-1\right) }-\frac{%
y\left( xy-x^{4}-1\right) \left( y+4x^{3}\right) }{x\left( xy+x^{4}-1\right)
^{2}}\right) -
\end{equation*}%
\begin{equation}
-\eta \left( \frac{\left( xy-x^{4}-1\right) }{x\left( xy+x^{4}-1\right) }+%
\frac{y}{\left( xy+x^{4}-1\right) }-\frac{y\left( xy-x^{4}-1\right) }{\left(
xy+x^{4}-1\right) ^{2}}\right) =0.
\end{equation}%
but in this case we have not found any solution, i.e. eq. (\ref{pato1}) does
not admit symmetries. Nevertheless, one always may try to find, as if by
magic, any c.v. that brings to find a simpler ode.
\end{proof}

\hrulefill

In this example we would like to show that there are some odes which are
very intractable. Our pedestrian method does not work in this case, we have
not been able to find any particular solution. The Lie method does not work,
i.e. this ode does not admit any symmetry and the theoretical methods do not
work either. At this time we do not know how to solve it.

\begin{example}
Try to solve the Abel ode
\begin{equation}
y^{\prime }=-Ax^{2}y^{3}-By^{2}-\frac{y}{x},  \label{pato2}
\end{equation}%
where $A$ and $B$ are dimensional constant.
\end{example}

\begin{proof}[Solution]
In this occasion we already have the dimensional constants%
\begin{equation}
\left[ A\right] =x^{-3}y^{-2},\quad \left[ B\right] =y^{-1}x^{-1},
\end{equation}%
in such a way that all the terms of the equation has dimensions of $\frac{y}{%
x}.$ We check if one of the dimensional relationship induces any particular
solution finding that this is not the case. Therefore we go next to look for
any trivial combination between them but we are not able to find any
particular solution. Nevertheless we try to obtain any result with the
following c.v.:
\begin{equation}
\left( t=\frac{B^{2}}{Ax},\quad u(t)=\frac{1}{Byx}\right) \Longrightarrow
\left( x=\frac{B^{2}}{At},\quad y=\frac{At}{B^{3}u(t)}\right)
\end{equation}%
which bring us to obtain the next ode:
\begin{equation}
t^{2}u^{\prime }u+tu+1=0,
\end{equation}%
but unfortunately we have not advanced.

Another try is the following one:
\begin{equation}
\left( t=\frac{1}{Byx},\quad u(t)=\frac{B^{2}}{Ax}\right) \Longrightarrow
\left( x=\frac{B^{2}}{Au(t)},\quad y=\frac{Au(t)}{B^{3}t}\right)
\end{equation}%
and hence:
\begin{equation}
tu^{2}+u^{\prime }\left( 1+ut\right) =0,
\end{equation}%
but as we supposed these attemps do not simplify our ode.

Now we change the strategy and we are going to suppose that the ode has
dimension of $y.$ In this case we need to introduce the following
dimensional constants in order to make eq. (\ref{pato2}) verify the p.d.h.
\begin{equation}
ay^{\prime }=-bx^{2}y^{3}-cy^{2}-a\frac{y}{x},
\end{equation}%
where $\left[ a\right] =x,\left[ b\right] =y^{-2}x^{-2}$ and $\left[ c\right]
=y^{-1},$ which bring us to the following c.v.
\begin{equation}
\left( t=x,\quad u(t)=\frac{1}{y}\right) \Longrightarrow \left( x=t,\quad y=%
\frac{1}{u(t)}\right)  \label{cvpato2}
\end{equation}%
therefore we obtain this new ode
\begin{equation}
u^{\prime }ut-2u^{2}-t^{5}-t^{2}u=0.
\end{equation}%
but as in the above tactic we have not advanced. This c.v. is precisely the
suggested one by the theoretic methods.

Following the Lie method we have to solve the pde:%
\begin{equation*}
\eta _{x}+({\eta }_{y}-{\xi }_{x})\left( -Ax^{2}y^{3}-By^{2}-\frac{y}{x}%
\right) -{\xi }_{y}\left( -Ax^{2}y^{3}-By^{2}-\frac{y}{x}\right) ^{2}-
\end{equation*}%
\begin{equation}
-\xi \left( \dfrac{1}{x^{2}}y-2Axy^{3}\right) -\eta \left( -2By-\dfrac{1}{x}%
-3Ax^{2}y^{2}\right) =0,
\end{equation}%
but this ode does not admit any symmetry.

To end, we would like to show that the theoretical method does not work
either. For this purpose we follow step by step the method beginning with a
generic Abel ode of first order written as follows:
\begin{equation}
y^{\prime }=f_{3}y^{3}+f_{2}y^{2}+f_{1}y+f_{0},
\end{equation}%
where in our case:
\begin{equation}
f_{3}=-x^{2},\quad f_{2}=-1,\quad f_{1}=-\frac{1}{x},\quad f_{0}=0.
\end{equation}

The c.v. suggested by the theoretical method is the following one
\begin{equation}
\left( t=x,\quad u(t)=\frac{1}{y}\right) \Longrightarrow \left( x=t,\quad y=%
\frac{1}{u(t)}\right) ,
\end{equation}%
(note that this c.v. is the same than the suggested one by D.A. (\ref%
{cvpato2})). In this way our ode is now rewritten as:
\begin{equation}
u^{\prime }u=h_{2}u^{2}+h_{1}u+h_{0},
\end{equation}%
where
\begin{equation}
h_{2}=-f_{1},\quad h_{1}=-f_{2},\quad h_{0}=-f_{3},
\end{equation}%
finding therefore that in our case we have:
\begin{equation}
u^{\prime }u=\frac{u^{2}}{t}+u+t^{2},
\end{equation}%
which is an Abel of second order (this ode has no solution). To try to find
a solution of this ode we make the following c.v.
\begin{equation}
\left( r=t,\quad s(r)=u(t)E\right) \Longrightarrow \left( t=r,\quad u(t)=%
\frac{s(r)}{E}\right) ,
\end{equation}%
obtaining this new ode:
\begin{equation}
ss^{\prime }=F_{1}s+F_{0},
\end{equation}%
where
\begin{equation}
E=\exp (-\int h_{2}),\quad F_{1}=h_{1}E,\quad F_{0}=h_{0}E^{2},
\end{equation}%
hence
\begin{equation}
ss^{\prime }=\frac{s}{r}+1,
\end{equation}%
but unfortunately we do not know how to solve this apparently simple ode.

As we have seen this ode seems very pathological since none of the followed
tactics have helped us to obtain any solution.
\end{proof}

\hrulefill

\section{Conclusions and discussion.}

We have seen how writing the odes in such a way that they verify the pdh
i.e. introducing dimensional constants, we can obtain in a trivial way c.v.
that bring us to obtain simpler ode than the original and therefore their
integration is immediate. Furthermore, we have tried to show that these c.v.
are not obtained as if by magic but that they correspond to invariant
solutions or to particular solutions and therefore they are generated by the
symmetries that admit the ode.

Nevertheless, the D.A. has \textbf{strong limitations}. For example D.A. is
unable (at least at this time we do not know how to do it) to solve the
following simple linear ode
\begin{equation}
y^{\prime }=\left( x^{3}+\frac{1}{x}+3\right) y+\left( 3x^{2}-\frac{1}{x^{2}}%
\right) ,
\end{equation}%
for this reason one must not put all his confidence in \ this
\textquotedblleft tactic\textquotedblright . This is one of the greater
inconvenience that presents the proposed method. But as we have noticed in
the introduction, we think that our pedestrian method continues having
validity at least when one is studying ode derived from engineering problems
or physical problems etc... where, as it is supposed, such odes must verify
the pdh in such a way that for example the ode $y^{\prime }=x+1,$ lacks of
any sense (physical sense, since we cannot add a number to a physical
quantity).

Nevertheless, and in spite of the limitations that we have not avoided to
show, we continued believing in the kindness (goodness) of the method and
that it can be applied to obtain solutions (at least particular solutions
and in concrete invariant solutions) to more complicated equations like the
following ones:%
\begin{eqnarray}
y^{\prime \prime } &=&\dfrac{y^{\prime 2}}{y}-ay^{2},\qquad \left[ a\right]
=y^{-1}x^{-2}\Longrightarrow y=\dfrac{1}{x^{2}}, \\
y^{\prime \prime \prime } &=&\dfrac{a}{y^{3}},\qquad \qquad \left[ a\right]
=y^{4}x^{-3}\Longrightarrow y=x^{3/4}, \\
y^{\prime \prime \prime } &=&-ayy^{\prime \prime },\qquad \qquad \left[ a%
\right] =y^{-1}x^{-1}\Longrightarrow y=x^{-1}, \\
y^{\prime \prime } &=&\dfrac{y^{\prime }}{x}+a\dfrac{3}{2}\dfrac{y^{2}}{x^{3}%
},\qquad \left[ a\right] =y^{-1}x\Longrightarrow y=x, \\
y^{\prime \prime \prime } &=&\dfrac{\left( y^{\prime \prime }\right) ^{2}}{%
y^{\prime }\left( 1+ay^{\prime }\right) },\qquad \left[ a\right]
=y^{-1}x\Longrightarrow y=x,
\end{eqnarray}%
but these are questions that we will approach in a forthcoming paper.

\begin{acknowledgement}
I would wish to thank to Prof. T. Harko for sending me his work on Abel
odes, to Prof. J. L. Caram\'{e}s to helping me to look for references and to
J. Aceves for translating this paper into English.
\end{acknowledgement}

\end{document}